\documentclass[prl, twocolumn, aps, 10pt]{revtex4-2}

\usepackage{amsmath,amssymb}
\usepackage{graphics,graphicx}
\usepackage{psfrag}
\usepackage{color}

\newcommand{\udg}{\affiliation{Departamento de Fisica, CUCEI, Universidad de Guadalajara, Guadalajara, Jalisco, Mexico}}

\begin{document}

 \title{Separate length scale for coarsening and for fractal formation by persistent sites}
%  in a model of global spin exchange dynamics. }
 \author{Dalia Hernandez  }
 \udg
 \author{Soham Biswas}
  \email{soham.biswas@academicos.udg.mx}
  \udg

\begin{abstract}
We present the first example where length scale for the growth of ordered regions and the correlation length for the two point correlations of persistent sites scale differently with time. We do so by studying a global spin exchange dynamics in one dimension where a selected spin interacts with its two nearest domains. We found domain growth exponent $z=2.47\pm 0.03$ and the persistence exponent $\theta=0.445\pm0.002$, making $z\theta > d=1$. Unlike any previous study, we found correlation length of two point correlation of persistent sites grows in a power law with exponent $\zeta=1.00\pm 0.03$ by studying the fractal structure created by the persistent sites at the different stages of the dynamics and shown that fractal dimension is not related with any growth exponents. 
\end{abstract}

\maketitle

 In last few decades physicists have studied several models for the opinion formation and segregation in a society \cite{Stau,Fortu,BooSen} that includes the models of non equilibrium spin flip dynamics \cite{Fortu} and spin exchange dynamics \cite{Stau,Mey}.
The change of opinion of the agents due to their surroundings has been modeled by spin flips \cite{Fortu,Weid}. On the other hand spin exchange dynamics have been studied when a couple of agents exchange their position for getting a better living environment of their choice \cite{Stau,Sche}.

These models of opinion dynamics not only enlighten the understanding of the collective behaviour of human society, but also give opportunity to investigate how such models can be classified into different dynamical classes based on their critical dynamical behaviour. One of the point of interest is the growth of the length scale of ordered regions with time when the system organises itself, starting from a random initial configuration. This is often studied in form of the growth of the average domain sizes $d_s\sim t^{1/z}$, where $z$ is the  ordering exponent or domain growth exponent \cite{Hohe,Krapi}.

Another point of interest is the persistence probability $P(t)$ (for spin dynamics, it is the probability that a spin will not change its orientation until time $t$) which is non-markovian by nature. $P(t)$ decays in a power law with time for several non-equilibrium physical phenomena \cite{Saty}, $P(t)\sim t^{-\theta}$ where $\theta$ is a non-trivial exponent known as persistence exponent which appeared to be unrelated to any previously known dynamical exponents. However, theory of finite size scaling of persistence probability connects the exponents $\theta$ and $z$, due to the underlying assumption that the length scale of ordered regions and the correlation length for two point correlation of persistent sites scale identically with time. These two exponents are related through  $z\theta= d-d_f $ where $d_f$ is the fractal dimension of the fractal structure formed by the persistent sites at late time \cite{MScal,Mfrac}. This happens as long as $z\theta < d$ \cite{MScal}. This condition has been seen to be true for all the previous cases where finite size scaling of persistence probability has been studied successfully \cite{Mmod1,Mmod2,Mmod3,Mmod4,BNNI}, even for the cases where the system orders without the presence of an conventional domain growth \cite{Mmod2, BNNI}. For $z\theta > d$, it has been  observed and argued that persistent sites no longer form a fractal \cite{Mmod5}.

The main purpose of this paper is to present the first example where the length scale for domain growth  and the correlation length for the two point correlations of persistent sites scale differently with time. By exploring the fractal structure created by persistent sites we also show that the fractal dimension is not necessarily related with the growth exponents. 

As candidate, we introduce a global spin exchange dynamics in 1-d where spin exchange happens under the influence of the size of two neighbouring domains,  modelling social segregation under social pressure. Psychological/social pressure felt by an individual is expected to be proportional to the number of similar/dissimilar people at the immediate neighbourhood of that person. An individual with similar orientation as that of the larger domain in its immidiate neighbourhood is considered to be happy. The dynamical rule of domain size dependence was first proposed for spin flips to model the change of opinion under social pressure for binary opinion dynamics \cite{SP}. These spin dynamics \cite{SP1} and the equivalent walker picture \cite{SP2} have been studied extensively in recent times.

We study the dynamics on one dimensional Ising spin system of size $N$, starting from a  random initial configuration. At each update, two  spins are selected at random. To mimic the domain size dependent dynamics, we define the following Hamiltonian [Eq. \ref{HSp}] for the system that will give the measure of energy of the selected spins ($s=\pm1$)
 \begin{equation} \label{HSp}
         H=-\sum_j s_j.\left(\sum_i s_i \right)
    \end{equation}
 index $i$ run over the spins of the two nearest domains at the both sides of the selected spin $j$. For the change of total energy of two selected spins $\Delta E<0$, these two spins deterministically swap their positions. For $\Delta E = 0$, they swap positions with probability $1/2$. Spin exchange happens only when two selected spins are of opposite orientation and at the boundary of two domains. Otherwise one Monte Carlo update is gone. One Monte Carlo time step is over after $N$ updates. 
 Starting from a random initial configuration the system reaches its minimum energy state, where only two domains remain. Periodic boundary condition is used and the results have been averaged over $ 10^3$ number of initial random configurations.  

\begin{figure}[ht]
\centering
\includegraphics[width=0.435\textwidth]{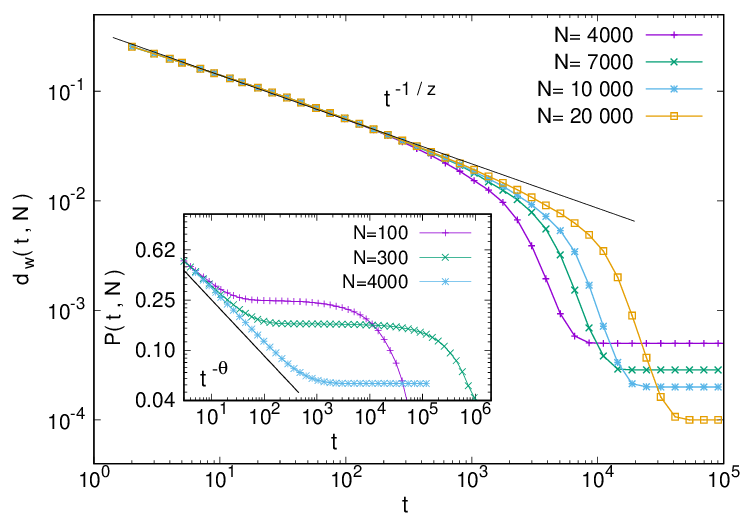}
\caption{ Main plot shows the power law decay of fraction of domain walls $d_w(t, N)$  until it reaches its minimum value $2/N$. Inset shows the decay of persistence probability $P(t,N)$ at the different stages of the dynamics.}
\label{FigDW}
     \end{figure}

Figure \ref{FigDW}, shows the power law decay for the fraction of domain walls $d_w\sim t^{-1/z}$ with $z=2.47\pm 0.03$. After the power law regime, $d_w$ decays exponentially to the saturation value $2/N$. 
Inset shows the decay of the persistence probability for different system sizes for large range of time. Persistence probability decays in a power law with exponent $ \theta = 0.445\pm0.002 $ to a metastable state, remains there for some time and then decays exponentially to zero. For computational limitations smaller system sizes are used to present the decay of persistence to zero. System keeps evolving even after reaching the ordered state that contains only two domains. Pair of spins of opposite orientations, located at different domain boundaries, keep exchanging their positions with probability $1/2$ which makes persistence decay to zero.

The values of $z$ and $\theta$ show $z \theta \nless d=1$. This immediately indicates that growth of correlation length for two point correlations of persistent sites must scale differently with time for the persistent sites to form a fractal structure. If this correlation length grows in a power law, it will be associated with a new growth exponent that is different from $z$.

Two-point correlation of the persistent sites $C(r,t)$ can be defined as the probability that site $i + r$ is persistent given that site $i$ is persistent (averaged over $i$) [App-A]
\begin{equation} \label{CrCalculo}
C(r,t)= \dfrac{\langle \rho_{i,t}\rho_{i+r,t}\rangle_i}{  \langle \rho_{i,t}\rangle_i }
\end{equation}
$\rho_{i,t}$ defines the density of the persistent sites. $\rho_{i,t}=1$ if site $i$ is persistent at time $t$ and $\rho_{i,t}=0$ if the site is not. $\langle \cdot \rangle_i$ is the average over the lattice and $\langle \rho_{i,t}\rangle_i = P(t)$ \cite{Mfrac}.

\begin{figure}[ht]
\centering
\includegraphics[width=.44\textwidth]{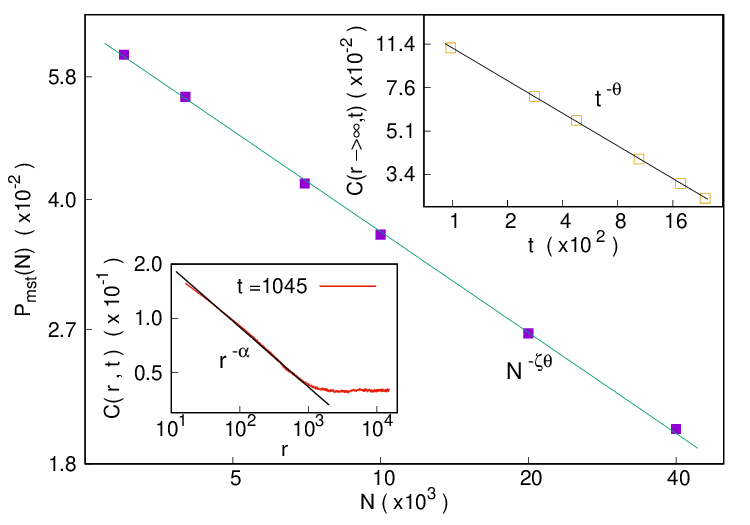}
\caption{Main plot shows the power law decay of $P_{mst}$ for different system sizes. Top inset shows the power law decay of $C(r\rightarrow \infty,t)$ in time with $\theta \approx 0.44$. Bottom inset shows the power law decay of $C(r,t)$ with $\alpha= 0.331\pm 0.002$.}
\label{PerCorrSatfig}
     \end{figure}

At any given time upto the metastable state  
\begin{equation} \label{CrPow}
    C(r)\sim r^{-\alpha} \quad\forall ~~ r ~~ with ~~ 1 \ll r \ll \xi(t)
\end{equation}
  indicating a strong spatial correlation between the persistent sites [lower inset Fig \ref{PerCorrSatfig}]. Correlation length $\xi$ separates the correlated and uncorrelated regions.

On the other hand, for large separations, persistent sites are uncorrelated. Hence two point correlator [Fig \ref{PerCorrSatfig}]
\begin{equation} \label{CRP}
    C(r,t)=P(t)\sim t^{-\theta} \quad\forall \quad r \gg \xi(t)
\end{equation}
  Over time, correlation length $\xi$ grows in a power law as 
\begin{equation}    \label{divxi}
    \xi(t)\sim t^{1/\zeta}
\end{equation}
where $\zeta$ is the growth exponent, different from the domain growth exponent $z$ and $\zeta \theta$ must be less than $d=1$. 

For the power law decay of $P(t)$, we can write $P(t)\sim t^{-\theta} \sim \xi(t)^{- \zeta \theta}$, using Eq. \ref{divxi}. At the metastable state, $\xi \rightarrow N$ and hence the saturation value of the persistence probability at the metastable state 
\begin{equation}\label{pmst}
P_{mst}(N)\sim N^{-\zeta\theta}
\end{equation}
Main plot of Fig.\ref{PerCorrSatfig} shows the decay of  $P_{mst}(N)$
with $\zeta\theta \simeq 0.44$ concluding $\zeta \approx 1.0$ that is different from $z \simeq 2.47$.

Power law decay of $C(r)$ for $r \ll \xi(t)$ indicates the formation of self-similar fractals by the persistent sites \cite{Mfrac}. This can be seen from the `Box-counting' method of computing the fractal dimension \cite{BC}. Let $M_t(s)$ be the average number of persistent sites in a box of length $s$ (discarding empty boxes) at time $t$. $M_t(s) \sim s^{d_f}$ where $d_f$ is the fractal dimension of the underlying fractal structure. $M_t(s)$ is related with $C(r,t)$ by 
\begin{equation}
\label{boxint}
 M_t(s)=\int_0^s C(r,t)dr
\end{equation}
 Using Eq. \ref{CrPow} and \ref{boxint}, we get $d_f=1-\alpha$. 
 $d_f$ can also be measured by computing the number of boxes $n_t(s)$ with at least one persistent site inside, at time $t$. $n_t(s) \sim s^{-d_f}$, for $d_f$ to be the fractal dimension. 

\begin{figure}[ht]
\centering
\includegraphics[width=.44\textwidth]{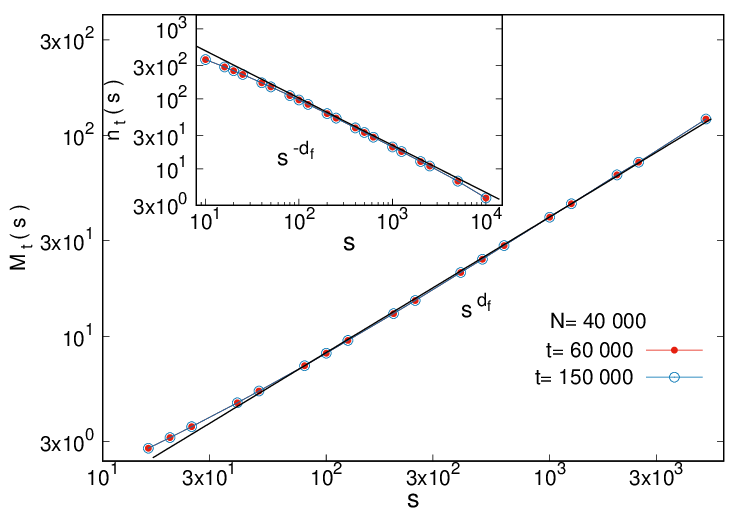}
\caption{ Main plot shows $M_t(s) ~ vs ~ s$. Inset shows computation of  $n_t(s)$ using the traditional Box-counting method. }
\label{Boxfig}
     \end{figure}

 For $r\gg\xi$ using Eq. \ref{CRP} and \ref{boxint}, we get $ M_t(s)= sP(t)$. 
 We measure $d_f$ at the metastable state where $\xi \rightarrow N$ making 
 $r \ll \xi$. From both, $M_t(s)$ and $n_t(s)$, we obtain $d_f= 0.675\pm 0.001$ [Fig. \ref{Boxfig}]. 

\begin{figure}[ht]
\centering
\includegraphics[width=.44\textwidth]{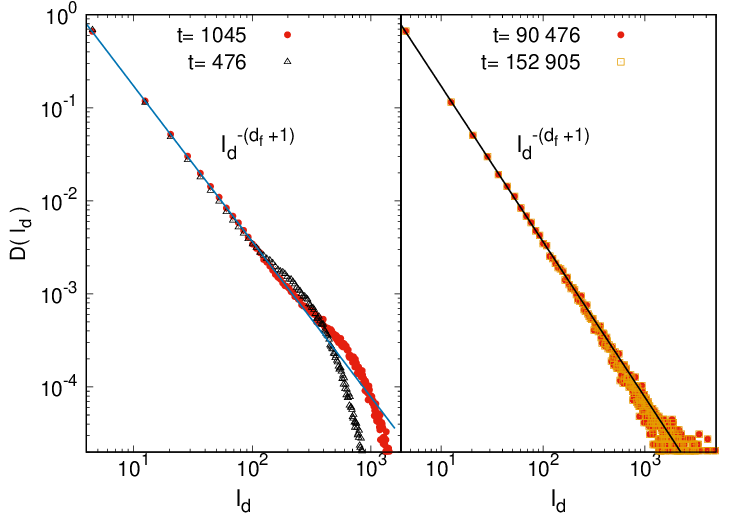}
\caption{Left plot shows $D(l_d)$ at earlier time when  $P(t,N)\sim t^{-\theta}$ ($\xi \ll N$), while the right graph shows $D(l_d)$ at late time when $P(t)$ is at the metastable state ($\xi \sim N$).}
\label{DistNonPer}
\end{figure}

Fractal dimension can also be obtained from the distribution of $l_{d}$: the distances between two nearest domains of persistent sites, if the  sizes of the persistent domains are small enough [Appendix-B]. The distribution function decays in a power law with exponent $d_f+1$ , for $d_f$ to be the fractal dimension \cite{Dist} for the fractal created in a 1-d ring. We computed $d_f$ at different times. Distribution function $D(l_d)$ is normalised with the total number of intervals between two persistent domains that removes the  time dependence [Fig. \ref{DistNonPer}]. Fractal formation happens at early stage of the dynamics and the fractal dimension $d_f$ remain same throughout the dynamics until the metastable state. $D(l_d)$ has a power law tail with exponent $1.67 \pm 0.003$ at any stage of the evolution concluding the fractal dimension $d_f=0.67\pm0.003$.

We obtain similar values of $d_f$ from different methods like `Box-counting', distribution function and two point correlation ($d_f=1-\alpha$), concluding $d_f=0.67\pm0.005$.

\begin{figure}[ht]
\centering
\includegraphics[width=.44\textwidth]{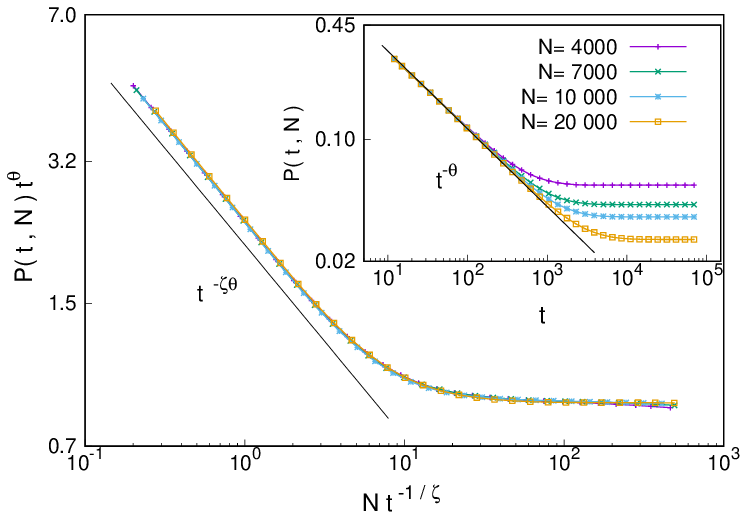}
\caption{ Data collapse of $P(t,N)$ for different system sizes $N$ (eq. \ref{DCPerEq}). Inset shows the raw data.}
\label{DCperfig}
     \end{figure}

 After understanding the dynamical behaviour of the system at early and late time, we can write the finite size scaling behaviour of $P(t)$. Combining the behaviour at short ($P(t) \sim t^{-\theta}$) and long time [Eq. \ref{pmst}] we write 
\begin{equation} \label{DCPerEq}
   P(t,N) \sim t^{-\theta}f(N/t^{1/\zeta})
\end{equation}
 $f(x)\sim x^{-\zeta\theta}$ for $x<<1$ (large $t$) and $f(x)$ is constant for $x \gg 1$ (small $t$).  $\zeta$ ($ \neq z$) is defined in Eq. \ref{divxi}. 
Figure \ref{DCperfig} shows the data collapse of the persistence probability for different system sizes obtained by choosing $\zeta=1.02\pm 0.01$ and $\theta=0.443\pm 0.003$. The values of the exponents are consistent with the previously obtained ones. 
       
One can also write the dynamical scaling form for $C(r,t)$ by summarising the behaviour for $r \ll \xi$ [Eq. \ref{CrPow}] and $r\gg \xi$ [Eq. \ref{CRP}]. Later we will see, for $r \sim \xi$, the behaviour of $C(r,t)$ is more complex. However that does not prevent us from writing the scaling behaviour of $C(r,t)$. A subtle time dependence of $C(r,t)$ is also present at the regime of power law decay ($r \ll \xi$) which can also be characterised by the scaling behaviour 
\begin{equation} \label{corsc}
 C(r,t)\sim P(t)g(r/\xi(t))\sim P(t)g(r/t^{1/\zeta})
\end{equation}
 $g(x)\sim x^{-\alpha}$ for $x\ll1$ ($r \ll \xi$) and $g(x)$ is a constant for $x\gg1$. For $x\ll1$ Eq. \ref{corsc} becomes 
 \begin{equation} \label{CorrExp}
 C(r,t) \sim P(t) r^{-\alpha}t^{\alpha/\zeta} \sim r^{-\alpha}t^{\alpha/\zeta-\theta}
\end{equation}
Using the previously obtained values of $\alpha$, $\theta$ and $\zeta$ we can predict that for a fixed $r (\ll \xi)$, $C(r,t) \sim r^{-0.331}t^{-0.105}$.

\begin{figure}[ht]
\centering
\includegraphics[width=.44\textwidth]{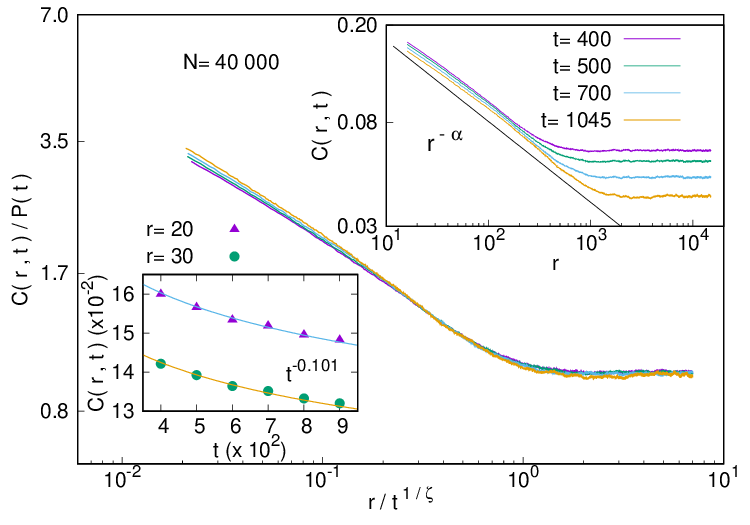}
\caption{Main plot shows the data collapse of $C(r,t)$.  Top inset presents the raw data of $C(r,t)$. Bottom inset shows the time dependence of $C(r,t)$ for $r<<\xi$ in a normal plot.} %\pm 0.002
\label{col_cor}
\end{figure}

 The existence of this time dependence is shown in the bottom inset of Fig.\ref{col_cor} for specific values of $r (<<\xi)$ using raw data. We obtained  $C(r,t) \sim r^{-0.331}t^{-0.101}$ which is close to the predicted value.
 Top inset of Fig.\ref{col_cor} shows the raw data for $C(r,t)$ measured at different times during the power law decay of persistence probability. Finally, in the main plot of Fig.\ref{col_cor} we show the data collapse for  $C(r,t)$ obtained independently using $\zeta=0.98 \pm 0.02 $ and the numerical values of $P(t)$ at time $t$.
 
If the behaviour of $C(r,t)$ for $r\ll \xi(t)$ and $r \gg \xi(t)$ coincide at $ r\sim\xi(t)$, one can write using Eq.\ref{CrPow} and Eq.\ref{CRP} : $\xi(t)^{-\alpha}\sim t^{-\theta} \Rightarrow \xi(t)\sim t^{\theta/\alpha}$. This expression along-with the fact $\xi\sim t^{1/\zeta}$ leads to the relation $\alpha=\zeta\theta$\cite{Mfrac}. However, we previously obtained $\alpha \approx 0.331 $ and $\zeta\theta\approx 0.44$ which confirms there exists a separate regime for the  behaviour of $C(r,t)$ when $ r\sim\xi(t)$. Note that, if $\alpha=\zeta\theta$, the time dependence at Eq.\ref{CorrExp} will not exist. Existence of this time dependence [lower inset of Fig.\ref{col_cor}] also conclude the existence of  a separate regime for $ r\sim\xi(t)$.

For finite sizes, it is complex to numerically characterise the behaviour of $C(r,t)$ at $ r\sim\xi(t)$. 
\begin{figure}[ht]
\centering
\includegraphics[width=.44\textwidth]{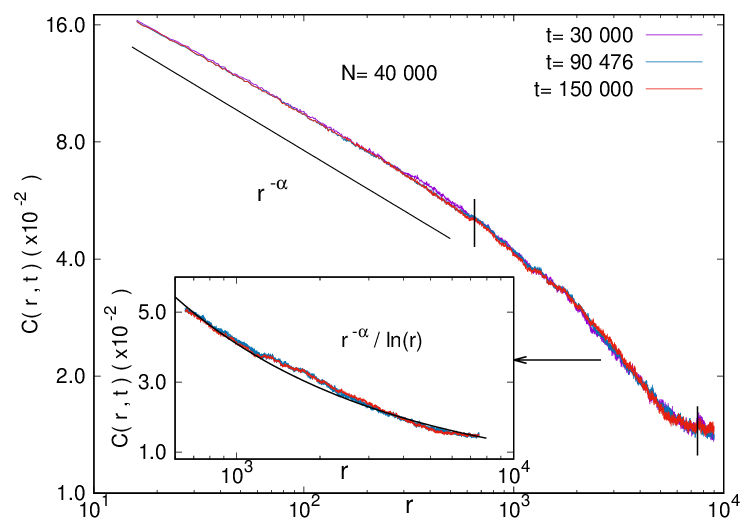}
\caption{$C(r,t)$ measured at late time once the power decay of $P(t)$ is over. Inset shows a zoom of $C(r,t)$ for a selected range of $r$ in a normal-log plot.}
\label{Corr40k}
     \end{figure}
When $\xi(t)$ is small (at short time), $C(r)$ saturates almost immediately after the power law decay making the regime $ r\sim\xi(t)$ invisible. Simulating $N=4\times10^4$ extensively \cite{Simulate} up to the metastable state (where $\xi(t)$ is large), we detect $C(r,t)$ has an intermediate decay behaviour with $r$, after the power law decay and before saturation [Fig.\ref{Corr40k}]. We observe two point correlation has a logarithmic correction over the power law decay as $C(r,t) \sim r^{-\alpha} / log(r)$ [Inset of Fig.\ref{Corr40k}] at this intermediate regime.

 Persistence probability has reached the metastable state with well defined fractal structure, by the time system reaches the ordered state that contains only two domains. The fractal structure slowly decay out mostly due to the random walk of two final domain walls, as spin exchange happens with probability $1/2$ between these two domain walls. Finally persistence probability reaches zero, erasing the existing underlying fractal. Persistence probability will remain in the metasteble state and will not go to zero if the spin exchange is not allowed for the zero change of total energy ($\Delta E=0$)[Appendix-B].

This work presents the first example to establish that the length scale for the growth of ordered regions and the correlation length for the two point correlations of persistent sites can scale differently with time. We have also shown that the fractal dimension of the underlying fractal is not related with both the growth exponents. 

The condition $z\theta > d$ does not guarantee the existence of the second length scale and the corresponding exponent $\zeta$. In that case, persistent sites will no longer form a fractal if there exists a single length scale in the system. This have been observed and studied for the non-conserved dynamics \cite{Mmod5}, where order parameter (magnetisation here) changes with time for a given configuration. The dynamics we study here is conserved which is the distinguishing and crucial ingredient we have in our model. Conserved dynamics with smart dynamical rule appears to be the key for the existence of the two different length scale in the system. Persistence properties for the conserved dynamics have never been studied before.

Studying persistence for the conserved dynamics would be a critical and important future direction of research. Global spin exchange dynamics (which is conserved) should be the point of interest as the system freezes for the local spin exchange dynamics (spin exchange happens within two nearest neighbours) for zero temperature \cite{kawasaki}. One should also take more careful approaches for using the scaling laws for the finite size scaling of   persistence in any future study. There are also several open questions like short time scaling for the `Box-counting' method, short and late time scaling of the domain size distribution that can lead to the future works for deeper understanding the model we studied. 

Global spin exchange dynamics was introduced and mostly studied to model the dynamics of social segregation that form large communities \cite{Sche}. From that point of view studying the global exchange dynamics with differnt dynamical rules in presence of the vacant sites (which represent empty houses for example) will be interesting and more realistic. 

Acknowledgements : We acknowledge Denis Boyer for useful discussions and the critical reading of a previous version of the manuscript. This work has been financially supported by the Conacyt project ``Ciencias de Frontera 2019'', number 10872.

\appendix 

\section{End Matter}

\textit{Appendix A : Computing two point correlation ::}
The two point correlation of the persistent sites $C(r,t)$ only increases its value when the sites i and  i+r are persistent, this means  $\rho_{i,t}\rho_{i+r,t}=1$ only if $\rho_{i,t}=\rho_{i+r,t}=1$, other cases  $\rho_{i,t}\rho_{i+r,t}=0$, including $\rho_{i,t}=\rho_{i+r,t}=0$.
A uncorrelated regime is when  $<\rho_{i,t}\rho_{i+r,t}>=<\rho_{i,t}><\rho_{i+r,t}>$. Hence $C(r,t)=<\rho_{i,t}>=P(t)$, when $\rho_{i,t}$ and $\rho_{i+r,t}$ are not correlated.

\begin{figure}[h]
\centering
\includegraphics[width=.44\textwidth]{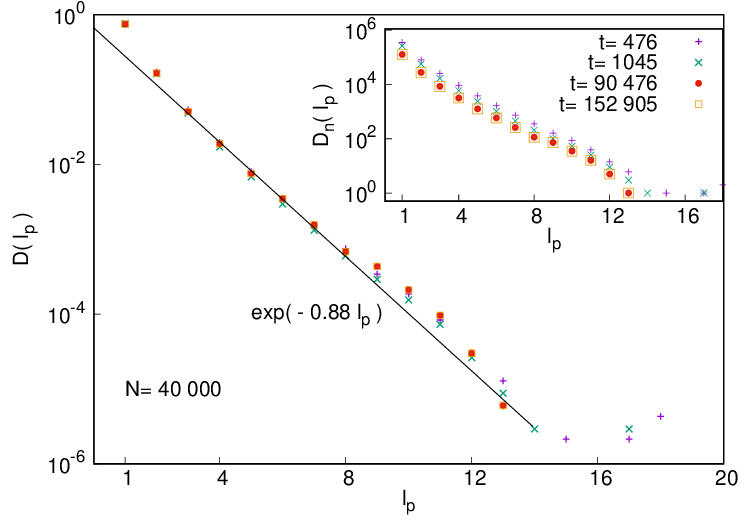}
\caption{ Probability and number distributions for the domain sizes of the persistent sites $l_p$.  Both the main plot and inset are in log normal. }
\label{DistPerfig}
     \end{figure}

  \textit{Appendix B : Distribution of persistent domains : }
Let us define $l_p$ as the domain sizes of the persistent sites.  Main plot of figure \ref{DistPerfig} shows the probability distributions $D(l_p)$ for different times. Inset shows the number distribution $D_n(l_p)$. These are exponentially decaying distributions with a small constant tail.

Figure \ref{DistPerfig} shows that the probability of finding a domain of persistent sites, of certain size, is almost similar at different time while the number of domains of persistent sites decreases with time.
     
\textit{Appendix C : The dynamics without stochastic exchanges ::}
Here we discuss the dynamics in the absence of stochastic spin exchanges.  In the paper we measure the change of energy $\Delta E$ for two randomly selected spins, if $\Delta E<0$ the spins definitely exchange positions and for $\Delta E=0$ the spins exchange with probability $1/2$. Now, we study the dynamics without stochastic spin exchanges in a similar way, the only difference is when $\Delta E=0$  we  deterministically  exchange spins  with probability 1 or with probability 0 maintaining the status-co. For all these three cases, the dynamical exponents of the measurables do not alter.

 At the ordered state of the spin system, the change of energy for exchanging the pair of spins of opposite orientations, located at different domain boundaries is equal zero. If spins exchange their positions with probability $1$, the lattice will rotate to the left or to the right,  depending on the randomly selected pair of spins and the persistence probability  will reach the zero value after the metastabke state.  If the status-co is maintained the persistence probability will saturate after the power law decay and will remain there.


\begin{thebibliography}{99}
\bibitem{Simulate} Simulating N=40000, for one random initial configuration  takes almost 3 days in a single core of Xeon E7. We average over 500 initial configurations.
\end{thebibliography}
\end{document}